\documentclass{article}
\usepackage{arxiv}
\usepackage{amsmath}
\usepackage{amsfonts}
\usepackage{hyperref}
\usepackage{url}
\usepackage{longtable}
\usepackage{booktabs}
\usepackage{array}
\usepackage{graphicx}
\usepackage{xspace}
\usepackage{xcolor}
\usepackage[labelfont=bf]{caption}
\usepackage{subcaption}
\usepackage{soul}

\newcommand{\fakeparagraphnospace}[1]{\noindent\textbf{#1}}

\newcommand\tlab[1]{\label{table:#1}}

\newcommand*{\defeq}{\stackrel{\text{def}}{=}}

\begin{document}

\title{Concentrated Liquidity with Leverage}

\author{
  Atis Elsts \\
  \texttt{atis@vertolabs.xyz} \\
  \And
  Krešimir Klas \\
  \texttt{kklas@kunalabs.io} 
}

\maketitle

\begin{abstract}
  Concentrated liquidity (CL) provisioning is a way how to improve the capital efficiency of Automated Market Makers (AMM).
  Allowing liquidity providers to use leverage is a step towards even higher capital efficiency.
  A number of Decentralized Finance (DeFi) protocols implement this technique in conjunction with overcollateralized lending.
  However, the properties of leveraged CL positions have not been formalized and  are poorly understood in practice.
  This article describes the principles of a leveraged CL provisioning protocol,
  formally models the notions of margin level, assets, and debt, 
  and proves that within this model, leveraged LP positions possess several properties that make them safe to use.
\end{abstract}

\section{Introduction}

Concentrated liquidity CL was first implemented by Uniswap v3~\cite{adams2021uniswap}, and has since been adopted by several other AMM protocols.
In addition to the capital efficiency provided by liquidity concentration, strategic liquidity providers (LP) may seek to take on debt to create their positions with borrowed assets.
Leveraged liquidity provisioning not only increases the income from fees, but can also protect the LP against price changes, particularly when single-sided leverage is used.
For instance, the LP may borrow the asset they deem less reliable, and in this way be insured against negative price action.
Furthermore, the combination of CL and leverage enables a separation between
  passive lenders, who provide single-sided assets to earn yield, and
  active liquidity providers, who create and manage leveraged positions, collect swap fees, and pay the costs for borrowing.

  At present, leveraged CL is offered by several DeFi protocols, including Panoptic~\cite{lambert2022panoptic}, Revert Lend~\cite{RL}, YLDR~\cite{YLDR}, and others.
  In particular, Kuna Labs~\cite{kunalabs} is developing a solution on the Sui blockchain, called Kai Leverage protocol.
  Its goals are to enhance capital efficiency in concentrated liquidity AMMs by utilizing leverage, and it is targeting the leading decentralized exchanges on Sui.

  Nevertheless, the properties and risks associated with leveraged CL are poorly formalized and not well understood in practice.
This article aims to address these shortcomings, by formalizing concepts related to leveraged CL, and to answer practitioner questions such as:
\begin{itemize}
\item If the margin level of a position is safe at two given prices, is it also safe everywhere in the range between these prices?
\item What kinds of operations are safe to perform with a leveraged position?
\item Can AMM spot price manipulation create undercollateralized positions and in this way create liquidations or cause the protocol to incur bad debt?
\end{itemize}

\newpage

\begin{table}[ht!]
  \caption{Symbols used.}
  \tlab{symbols}
\centering
\bgroup
\begin{tabular}{c|l}
\toprule
\textbf{Symbol} & \textbf{Description} \\
\midrule
    $X$ & Base token \\
    $Y$ & Quote token \\
    $x$ & Base token amount \\
    $y$ & Quote token amount \\
    $P$ & Price \\
    $p_a$ & Price range lower bound \\
    $p_b$ & Price range upper bound \\
    $S$ & Square root of price \\
    $V$ & Value \\
    $L$ & Liquidity \\
    $DL$ & Divergence loss (impermanent loss) \\
    $M$ & Margin level \\
    $\beta$ & Liquidation bonus \\
\bottomrule
\end{tabular}
\egroup
\end{table}


\section{Concepts and definitions}

\fakeparagraphnospace{CL pools and positions.}

A concentrated liquidity AMM features liquidity pools and liquidity positions.
An active pool contains one or more liquidity positions.
A position is fully defined by the triple $(L, p_a, p_b)$.
A position of type $(L, 0, \infty)$ is called a full-range position and is defined solely by the liquidity $L$.

A constant-product CL AMM follows the swap invariant $x_{virtual} \cdot y_{virtual} = L^2$.
The virtual amounts of assets $X$ and $Y$ can be translated into the real amounts $x_{pos}$ and $y_{pos}$ in the position using the price bounds and the current price $P$~\cite{liquiditymath}.
If $P \le p_a$, the position contains only $Y$, and if $P \ge p_b$ the position contains only $X$. Such a position is called \emph{out of range}. A position is \emph{in range} if $p_a < P < p_b$.

\fakeparagraphnospace{Value of a CL position.}

The value of a CL AMM position $(L, p_a, p_b)$ at price $P$ is:
\begin{equation}\label{eq:posval}
V_{pos}(P) =
\begin{cases} 
L \cdot \frac{\sqrt{p_b} - \sqrt{p_a}}{\sqrt{p_a} \cdot \sqrt{p_b}} \cdot P,  & \text{if } P \leq p_a \\
L \cdot (2 \sqrt{P} - \sqrt{p_a} - P/\sqrt{p_b}), & \text{if } p_a < P < p_b \\
L \cdot (\sqrt{p_b} - \sqrt{p_a}), & \text{if } P \geq p_b
\end{cases}
\end{equation}
To be clear, this expression 
does not include any accumulated fees that may also be present.

\fakeparagraphnospace{Initial value of assets and debts.}

The initial assets of a CL provider consist of two components:
\begin{enumerate}
\item Their own capital $(x_{user}, y_{user})$ with value $V_{user}$.
\item Borrowed capital $(x_{D}, y_{D})$ with value $V_{D}$.
\end{enumerate}
It follows that the initial value of the assets is:
\begin{equation}
  A(P_0) = (x_{user} + x_{D}) \cdot P_0 + y_{user} + y_{D}
\end{equation}

It is possible that only some of these assets are put in the CL position, leaving some of the $X$ or $Y$ as extra collateral $(x_C, y_C)$ with value $V_C$.
\begin{align}
  x_{C} &= x_{D} + x_{user} - x_{pos}(P_0) \\
  y_{C} &= y_{D} + y_{user} - y_{pos}(P_0)
\end{align}

Consequently, the initial value of assets can be rewritten as:
\begin{equation}
  A(P_0) = V_{pos}(P_0) + V_{C}(P_0)
\end{equation}

The initial value of the debt $(x_D, y_D)$ is defined as:
\begin{align}
  D(P_0) &=  x_{D} \cdot P_0 + y_{D} =\\
      & = \text{max}(x_{pos}(P_0) - x_{user}, 0) \cdot P_0 + \text{max}(y_{pos}(P_0) - y_{user}, 0)
\end{align}

For a leveraged position where both assets are borrowed, the initial value of the debt is equal to:
\begin{align}
  D(P_0) &= V_{pos}(P_0) - V_{user}(P_0)
\end{align}

\fakeparagraphnospace{Evolution of assets and debt.}

1. The value of the \textbf{debt} evolves as a linear function of the price, similar to the value of a cryptocurrency holder's portfolio:
\begin{align}
  D(P) &= x_D \cdot P + y_D
\end{align}

Moreover, in the special case where only $Y$ is borrowed, the debt is a constant function, unaffected by the price:
\begin{align}
  D_{y\_only}(P) &= D(P_0)
\end{align}

In another special case, when both assets are borrowed, the value of the debt is:
\begin{align}
  D_{both}(P) &= V_{hold}(P) - V_{user}(P) \,\text{,}
\end{align}
where $V_{hold}(P)$ is the value of the buy-and-hold portfolio at price $P$. The asset composition of this buy-and-hold portfolio is the same as the asset composition of the position at $P_0$.
\\

2. The value of the \textbf{assets} evolves non-linearly. It is equal to the sum of the position's value (Eq.~\ref{eq:posval}) and the extra collateral:
\begin{align}
  A(P) &= V_{pos}(P) + V_{C}(P)
\end{align}

\fakeparagraphnospace{Margin level and leverage.}

\begin{figure}[t]
    \centering
    \begin{subfigure}[b]{0.48\textwidth}
        \centering
        \includegraphics[width=\textwidth]{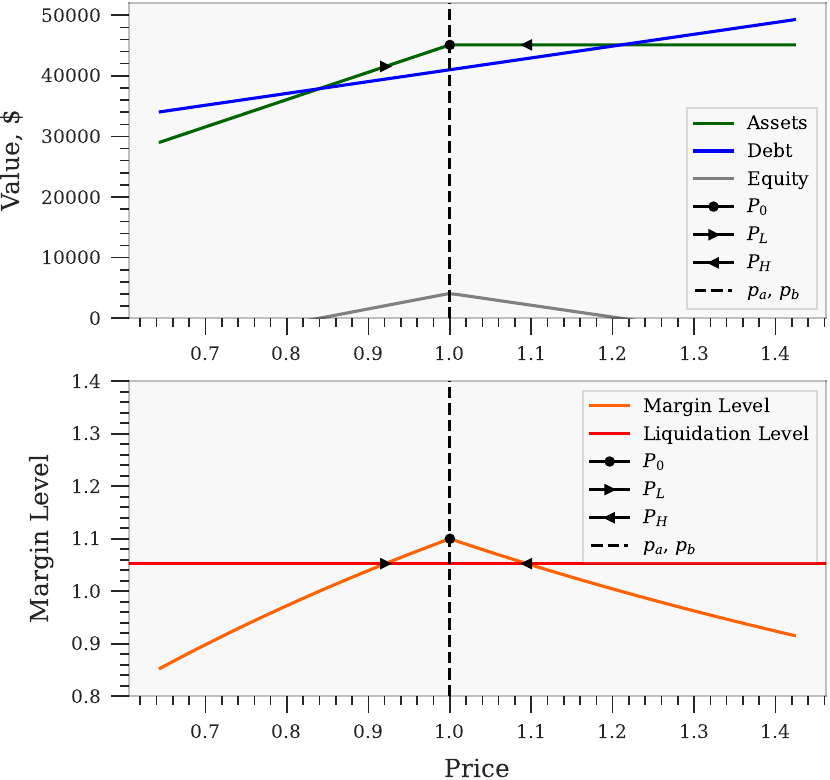}
        \caption{Stable pair, highly concentrated position}
        \label{fig:model1}
    \end{subfigure}
    \hfill
    \begin{subfigure}[b]{0.48\textwidth}
        \centering
        \includegraphics[width=\textwidth]{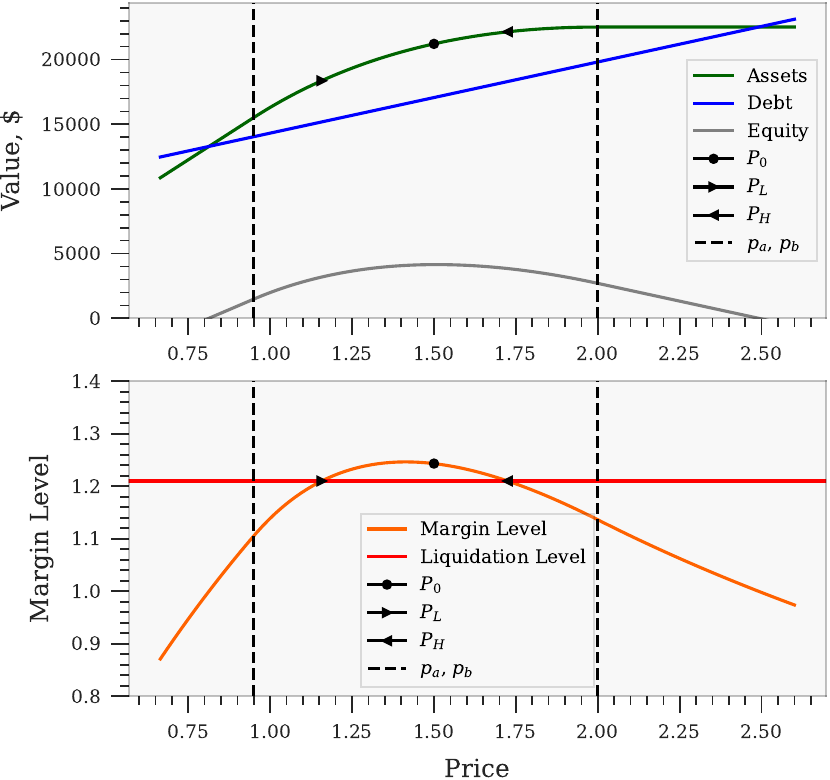}
        \caption{Volatile pair, less concentrated position}
        \label{fig:model2}
    \end{subfigure}
    \caption{Examples of margin level functions for stable and volatile pairs.}
    \label{fig:models}
\end{figure}

Let us define \textbf{margin level} $M$ as the relation between the assets and the debt:
\begin{align}\label{eq:ml}
  M(P) &= \frac{A(P)}{D(P)} = \\
  &= \frac{V_{pos}(P) + V_C(P)}{V_D(P)} = \\
  &= \frac{V_{pos}(P) + x_C \cdot P + y_C}{x_D \cdot P + y_D}
\end{align}
See Fig.~\ref{fig:models} for examples.

A position's margin level is related to its leverage factor through the following formula:
\begin{align}
&  \text{leverage}(P) = 1 + \frac{1}{M(P) - 1}
\end{align}
For instance, a margin level of $1.5$ corresponds to a leverage of $3$x.

\fakeparagraphnospace{Liquidations.}

If lending protocol is to remain solvent, it must ensure that the value of assets always exceeds the value of debt.
In practice, protocols do not allow margin levels to fall below predefined safe parameters and will initiate liquidations if they do.

A healthy position is defined as one with $M > M_L$, where $M_{L}$ is the liquidation threshold.
A liquidation mechanism attempts to repay part or all of position's debts in order to rise its margin level to another specific level $M_T > M_L$, while also paying a liquidation bonus proportional to the value repaid.
If that is not possible, 
the mechanism attempts to repay as much of the debt as possible, while still paying out the liquidation bonus first.

In practice, $M_L$, $M_T$, and the liquidation bonus proportion are parameters of the protocol and may vary for different asset pairs. However, all the thresholds are required to be strictly above 1.0, since margin level of 1.0 corresponds to an infinite leverage.

\fakeparagraphnospace{Other operations.}

Before a position can be liquidated, it must be withdrawn from the pool.
It is possible that withdrawing the assets and repaying that already increases the margin level above $M_L$, thereby avoiding liquidation.
This operation is called \emph{deleveraging}.
Formally, it decreases the value of both the assets and the debt by the same amount, $V_{\text{repaid}}$, thereby increasing the margin level:
\begin{align}
  M'(P) = \frac{A(P) - V_{\text{repaid}}}{D(P) - V_{\text{repaid}}}
\end{align}
Only fully deleveraged positions can be liquidated; that is, all assets must be withdrawn from the pool, and as much of the debt as possible must be repaid without swapping between $X$ and $Y$.

The user also may want to \emph{reduce} their position, i.e., decrease their exposure or otherwise to withdraw all or a portion of their assets while keeping the margin level intact.
A reduction mechanism implemented by the protocol allows users to perform this operation by reducing both their assets and debt by a factor of $k$:
\begin{align}
  M'(P) = \frac{(1 - k) A(P)}{(1 - k)D(P)}
\end{align}

\fakeparagraphnospace{Other remarks.}

Liquidity providers earn swap fees from traders and must pay borrow fees to the liquidity lenders. However, we assume that these fees are relatively small and can be ignored in short-term solvency calculations.

For long-term dynamics, we envision an automated bookkeeping process that collects the swap fees and uses them to repay the debt or compound the liquidity positions, effectively changing the structure of the debt. Analyzing such a mechanism goes beyond the scope of this article.

\section{Analysis}


\subsection{Properties of the system}

This article demonstrates the following properties of a leveraged CL protocol:
\begin{enumerate}
\item \textbf{In-interval safety:} if the margin level function $M(P)$ is not below a threshold $M_L$ at both ends of a price interval $[P_L, P_H]$, it is also not below the threshold for every point $P$ in the interval $P_L < p < P_H$.
\item \textbf{Computability of price bounds:} for a given threshold  $M_L$ and liquidity level $L$, there are two functions $PL(M_L)$ and $PH(M_L)$ that determine the prices $P_L$ and $P_H$ where the threshold  $M_L$ is reached.
\item \textbf{Computability of maximum safe liquidity:} the safe liquidity level for a given price interval $[P_L, P_H]$ and threshold $M_L$ can be numerically computed using binary search.
  \item \textbf{Safe reductions:} reduction of a healthy position ($M > M_L$) does not change its margin level.
  \item \textbf{Safe deleverages:} deleveraging of a positive-margin position ($M > 1.0$) increases its margin level.
  \item \textbf{Safe liquidations:} liquidation of a position above a critical margin level ($M > M_C$) both makes its remainder healthy and pays out a liquidation bonus in full.
\item \textbf{Safety from price manipulation:} spot price manipulation in the AMM pool cannot decrease the value of a position.
\item \textbf{Connection with divergence loss:} As leverage increases, the shape of the margin level function approaches the shape of the divergence loss function.
\end{enumerate}

The following subsections analyze each of these properties in turn.

\subsection{In-interval safety}

\begin{figure}[t]
    \centering
    \includegraphics[width=0.6\textwidth]{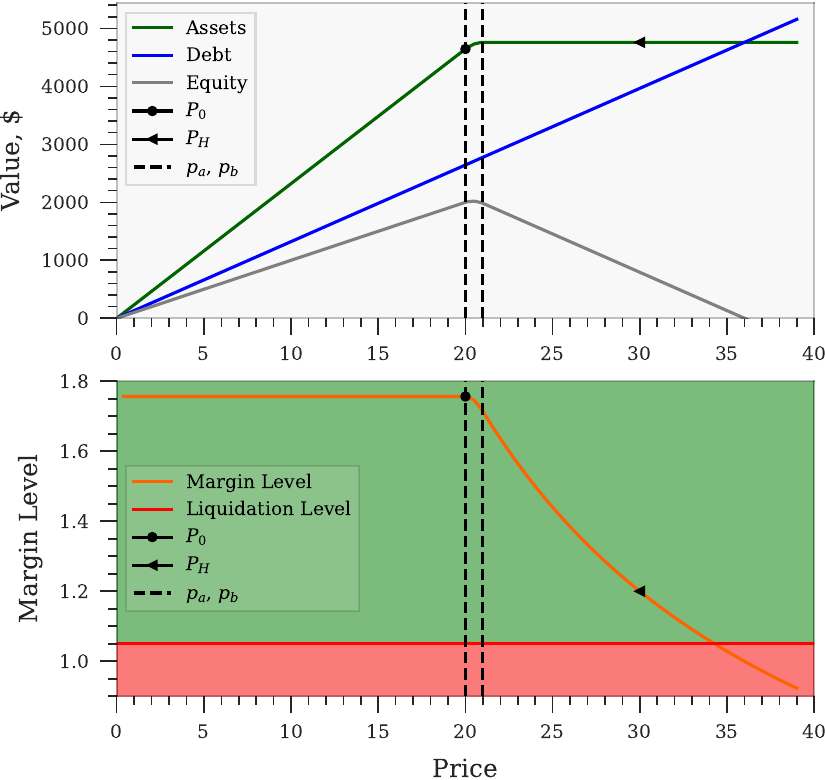}
    \caption{Example of in-interval safety: the whole range between two prices such as $P_0$ and $P_H$ are guaranteed to be safe from liquidation, because both $P_0$ and $P_H$ are safe. Margin levels safe from liquidation shown with green background, other levels with red background.}
    \label{fig:in-interval-safety}
\end{figure}

It is convenient to rely on the fact that if the margin level is safe at both ends of a price interval, it is guaranteed to remain at a safe level everywhere in the interval (Fig.~\ref{fig:in-interval-safety}).
As long as the price stays within this interval, the position cannot be liquidated.
More formally, if $M(P_L) \ge M_L$ and $M(P_H) \ge M_L$, then it should be the case that:
\begin{align}
&\forall P: P_{L} < P < P_{H} \Rightarrow M(P) \ge M_L
\end{align}
This property allows the protocol and the user to check the margin levels only at the interval's endpoints, rather than checking every point within the interval.

This property is true if the margin function does not have any local minima.
Such is the case for Uniswap-v3-like CL AMM, where the position's value function follows Eq.~\ref{eq:posval}.
For a proof of see the Appendix.

\subsection{Computability of price bounds}

In addition to computing the margin level at a given price, one may wish to solve the inverse problem:
computing the price targets $P_L$ and $P_H$ where a predefined margin level $M_L$ is reached.
Specifically, we are interested in this problem when $P_L < P_0 < P_H$, where $P_0$ is the initial price at which the position is deployed.

A typical margin function is shaped like an inverted parabola in its domain of definition ($P>0$).
However, several special cases exist where $M(P)$ can be monotonically increasing or decreasing.
In general, the following algorithm can be used:
\begin{enumerate}
\item Check $M(P_0)$. If it is below the threshold, terminate early: the position cannot be safely deployed.
\item Check $M(p_a)$:
  \begin{enumerate}
  \item If it is above the threshold, then $P_L \le p_a$ due to the in-interval safety property, and $P_L$ either is zero, or can be found by solving a simple linear equation.
  \item Otherwise, $p_a < P_L \le P$, and $P_L$ can be found using an expression given in the Appendix.
  \end{enumerate}
\item Check $M(p_b)$:
  \begin{enumerate}
  \item If it is above the threshold, then $P_H \ge p_b$ due to the in-interval safety property, and $P_H$ is either infinity, or can be found by solving a simple linear equation.
  \item Otherwise, $P \le P_H < p_b$, and $P_H$ can be found using an expression given in the Appendix.
  \end{enumerate}
\end{enumerate}
The expressions for in-range $P_L$ and $P_H$ were found using Python's \texttt{sympy} package.

\subsection{Computability of maximum safe liquidity}

The problem here is to compute the maximum liquidity level $L_{max}$ that is permissible, i.e., $M(P) > M_L$ for all prices $P$ in an interval $[P_L, P_H]$.

Due to the in-interval safety property, $L_{max} = \text{min}(L_{max}(P_L), L_{max}(P_H))$, where $L_{max}(P)$ is the value of $L$ at which $M(P) = M_L$.

The price levels $P_L$ and $P_H$ can be user- or protocol-configured parameters, expressed in terms of the initial price, for example, using a range factor $r$:
\begin{align}
  P_L &= P_0 / r\\
  P_H &= P_0 \! \cdot \! r
\end{align}

Exact solutions of the margin function in terms of the target liquidity level are possible and have been found.
However, they have a complicated, case-by-case nature depending on whether $P_L$ and $P_H$ are in range, as well as whether both $X$ and $Y$ or just one of those assets has been borrowed.
In practice, it may be more convenient to use an efficient search algorithm, such as binary search, to compute the solutions numerically.
The convergence of binary search to the right solution is guaranteed due to the absence of local minima in the margin level function; for a proof of the latter, see the Appendix.

\subsection{Safe reductions}

The protocol allows the users to reduce healthy positions, i.e., when $M(P) > M_L$.
By reducing a position, the user aims to gain some free assets while keeping the position's margin level intact.
The margin level of a position reduced by a factor $k$ does not change, provided $0 < k < 1$:
\begin{align}
  &  M'(P) = \frac{(1 - k) A(P)}{(1 - k) D(P)} = M(P)
\end{align}
Let's define:
\begin{align}
  V_{\text{removed}} &\defeq k A(P) \\
  V_{\text{repaid}} &\defeq k D(P) \\
  V_{\text{freed}} &\defeq V_{\text{removed}} - V_{\text{repaid}}
\end{align}
By assumption, $M(P) > M_L > 1.0$, and follows that $k A(P) > k D(P)$. Hence, $V_{\text{freed}}$ is strictly positive: the user is guaranteed to obtain some free assets after reducing a healthy position.

\subsection{Safe deleverages}

Deleveraging removes some assets from a position and uses them to repay the debt:
\begin{align}
  &  M'(P) = \frac{A(P) - V_{\text{repaid}}}{D(P) - V_{\text{repaid}}} ~ \text{,}
\end{align}
where $0 < V_{\text{repaid}} <= D(P)$.

The difference between deleveraging and liquidation is that deleveraging only repays the assets that are already in the position, i.e. without swapping them, and as such, it is not subject to price risks to the same extent as liquidations.

Assuming that $M(P) > 1.0$, then $A(P) > D(P)$. Let us introduce $\varepsilon \defeq A(P) - D(P)$
\begin{align}
  M(P) &= \frac{A(P)}{D(P)} = 1 + \frac{\varepsilon}{D(P)} \\
  M'(P) &= \frac{A(P) - V_{\text{repaid}}}{D(P) - V_{\text{repaid}}} = 1 + \frac{\varepsilon}{D(P) - V_{\text{repaid}}}
\end{align}
Since $D(P) > D(P) - V_{\text{repaid}}$, it follows that $M'(P) > M(P)$: deleveraging can only increase the margin level of a position if the assumption $M(P) > 1.0$ is met.

Note that this property is not true for positions with margin level at or below 1.0.

\subsection{Safe liquidations}

Liquidations convert some of the user's assets to fully or partially repay their debt.
The margin level after a liquidation changes in a similar way as it does after deleveraging, execept that there is an additional term representing the liquidatin bonus:
\begin{align}
  &  M'(P) = \frac{A(P) - V_{\text{repaid}} - V_{\text{bonus}}}{D(P) - V_{\text{repaid}}} ~ \text{,}
\end{align}
The liquidation bonus is proportional to the value repaid:
\begin{align}
  & V_{\text{bonus}} \defeq \beta V_{\text{repaid}}
\end{align}

The value $V_{\text{repaid}}$ depends on the initial margin level $M(P)$ of the position, the liquidation threshold $M_L$, the target threshold $M_T$, and the liquidation bonus proportion $\beta$.

There is a critical margin level threshold $M_C$, below which the debt cannot be fully liquidated. At the threshold $M_C$, the value repaid is equal to the value of the debt, and consumes all of the assets:
\begin{align}
  \begin{cases}
  A(P) = V_{\text{repaid}} + \beta V_{\text{repaid}} \\
  D(P) = V_{\text{repaid}}
  \end{cases}
\end{align}
It follows that:
\begin{align}
  M_C = \frac{(1 + \beta)V_{\text{repaid}}}{V_{\text{repaid}}} = 1 + \beta
\end{align} 

To ensure safe operation of the protocol, the values of $M_L$ and $M_T$ should be set so that $M_C < M_L < M_T$.
If that is done, and we define $V_{\text{repaid}}(P)$ in term of asssets --- as $k A(P)$ --- then the coefficient $k$ can be computed as:
\begin{align}
& k = 
  \begin{cases}
    0 \,\text{,}                                     & \text{if } M \ge M_L  \\
    \frac{M_T - M}{(M_T - (1 + \beta)) M} \,\text{,} & \text{if } M_L > M > M_C \\
    \frac{1}{1 + \beta} \,\text{,}                   & \text{if } M \le M_C
\end{cases}
\end{align}

Case-by-case:
\begin{enumerate}
\item If the margin level $M \ge M_L$, the liquidation does not happen.
\item If the margin level $M_L > M > M_C$, the debt is fully liquidated, and the new margin level $M'$ after the liquidation is:
\begin{align}
  &  M' = \frac{A(P) - k A(P) - k \beta A(P)}{D(P) - k A(P)} = \frac{1 - (1 + \beta) k}{\frac{1}{M} - k}
\end{align}
\item If the margin level $M = M_C$, the debt is fully liquidated, but so are assets.
\item If the margin level $M < M_C$, the assets are fully liquidated, but some debt remains. We assume that paying out the liquidation bonus in full is a priority. If the protocol sets the priorities differently, the formula of $k$ for this case should be changed.
\end{enumerate}

In the first case, the new margin level $M'$ is equal to $M$, and in the third and fourth cases it does not exist. Only the second case need to be analyzed to show that $M'$ is larger than $M$, i.e., the liquidation is safe. Algebraically, $M'$ can be shown to be equal to $M_T$, which by assumption is more than $M$.

Additionally:
\begin{itemize}
\item The value of $k$ sets a lower bound on what is required for safe operation.
  Protocols may choose to opt for more aggressive liquidations, where a higher proportion of the assets are liquidated at higher margin levels.
  For instance, the Kai Leverage protocol (Section~\ref{sec:practical}) fully liquidates a position if its margin level falls below $(M_L+M_C)/2$.
\item If the margin level is below $M_C$, the debt $D(P)$ cannot be repaid in full while also paying out the liquidation bonus.
  The protocol may prioritize liquidators in this case, and fix the value of the bonus equal to $\beta  D(P)$ to ensure sufficient incentives for liquidation.
  The protocol should strive avoid these situations, as they create bad debt for it.
\end{itemize}

\subsection{Safety from price manipulation}

It is considered easy to manipulate the spot price of an AMM for short periods of time.
If this could be used to reduce the margin level of a position below $M_L$, it could make healthy positions eligible for liquidation.
Moreover, if the level could be reduced below $1 + \beta$, the protocol could be forced to incur bad debt.
Therefore, it is crucial to demonstrate that the protocol is robust against spot price manipulation, as long as it has access to an accurate price from an oracle.

To prove this, first consider that spot price manipulation cannot change the value of an out-of-range position, since it does not alter its composition.
Without loss of generality, we can assume that (1) the position is in-range, and (2) the CL pool has non-zero liquidity (otherwise, all positions are out of range, which contradicts the first assumption).
Price manipulation in CL pools with non-zero liquidity can be modeled as a series of price manipulations in $xy=k$ AMM pools.
Therefore it is sufficient to prove this property for $xy=k$ AMMs.
Additionally, the presence of any swap fees makes the attack more costly for the perpetrator, so we will focus on the case without swap fees.

Consider a swap that exchanges some $X$ and $Y$ tokens.
Let us denote the changes in the AMM pool's balance by $\Delta x$ and $\Delta y$, where the signs of the deltas depend on the direction of the  swap.
The value of the position after the swap, as computed using the true (oracle) price $P$, is:
\begin{align}
  V'_{pos} = V_{pos} + \Delta V = V_{pos} + \Delta x \cdot P + \Delta y
\end{align}
The deltas can be computed from the liquidity $L$, the current price $P$ (assumed to match the oracle price) and the price after the swap $P'$ (see Eqs.~6.14 and 6.16 in~\cite{adams2021uniswap}):
\begin{align}
  \Delta x  = \frac{L}{\sqrt{P'}} - \frac{L}{\sqrt{P}} \\
  \Delta y  = L \sqrt{P'} - L \sqrt{P}
\end{align}
Let us compute the change in the position's value:
\begin{align}
  & \Delta V =  \Delta x \cdot P + \Delta y = \\
  &= (\frac{L}{\sqrt{P'}} - \frac{L}{\sqrt{P}}) P + L \sqrt{P'} - L \sqrt{P} = \\
  &= L \frac{P - 2 \sqrt{P'} \sqrt{P}  + P'}{\sqrt{P'}} = \\
  &= L \frac{(\sqrt{P} - \sqrt{P'})^2}{\sqrt{P'}}
\end{align}
The square in the numerator guarantees that $\Delta V$ is positive, since $L, P, P'$ are all positive.
We have thus proved that any price-manipulating swap, under the assumptions stated above, always increases the value of the assets in a position.
Liquidations of healthy positions cannot be forced in this way.

\subsection{Connection with divergence loss}

The relative divergence loss (DL, also called ``impermanent loss'' or ``rebalancing loss'') of a position is defined as:
\begin{align}\label{eq:dl}
  DL_{relative}(P) &= \frac{V_{pos}(P) - V_{hold}(P)}{V_{hold}(P)} = \\
  & = \frac{V_{pos}(P)}{V_{hold}(P)} - 1\label{eq:dl2}
\end{align}

The margin level function (Eq.~\ref{eq:ml}) when both assets are borrowed ($x_C = y_C = 0$) reduces to:
\begin{align}\label{eq:ml-reduced}
  &  M(P) = \frac{V_{pos}(P)}{V_{hold}(P) - V_{user}(P)}
\end{align}
The similarity between Eq.~\ref{eq:dl2} and Eq.~\ref{eq:ml-reduced} shows that as $V_{user} \rightarrow 0$, the margin level $M(P)$ $ \rightarrow DL(P) + 1$.

It follows that the losses of a leveraged concentrated liquidity LP that borrows both assets are the same
as the DL of the full position (when using buy-and-hold at the initial token proportions as the benchmark).

Relative to the starting capital of the LP, these losses are the same as the DL multiplied by the position's initial leverage factor.

\section{Practical Aspects}
\label{sec:practical}

This section is specific to the implementation in the Kai Leverage protocol, although many aspects are common with other implementations.

\subsection{Supply Pools}

Supply pools are single-sided liquidity pools used as a source of liquidity for leveraged positions.
Liquidity deposited in these pools earns interest as assets are borrowed for leveraged position creation.
The interest model is based on a piecewise curve that increases the interest rate as pool utilization increases.

It is worth noting that each supply pool has a separate interest model associated with each DEX pool over which the leveraged positions are created.
This means that supply pool liquidity providers can earn higher interest rates when their assets are utilized in leveraged positions with higher risk, e.g., more in the SUI/USDC pool than in the USDC/USDT pool.

\subsection{Position Creation}

For position creation, the following assertions are made:

\begin{enumerate}
\item The current pool price is within the position range.
  \begin{itemize}
  \item Formally: $p_a \le P_0 \le p_b$.
  \end{itemize}
\item The position's margin levels at prices \texttt{min\_liq\_start\_price\_delta} away from the current price (in both directions) are above $M_L$ (\texttt{liq\_margin} in the code).
  \begin{itemize}
  \item Formally: $\text{min}(M((1 - \Delta P) P_0), M((1 + \Delta P) P_0)) > M_L$.
  \end{itemize}
\item The position's initial margin is above \texttt{min\_init\_margin}.
  \begin{itemize}
  \item Formally: $M(P_0) > M_0$.
  \end{itemize}
\end{enumerate}

Assertion (3) is not strictly necessary, as the first two assertions already imply the maximum leverage, but it is useful in practice to have this check to limit the potential risk of high leverage.

There are two other constraints that are not part of the leverage model but are useful for limiting liquidation and ecosystem risk --- \texttt{max\_position\_l} and \texttt{max\_global\_l}.
The first one limits the maximum liquidity that can be used in a single position, and the second one limits the total liquidity that can be used in all positions across a single pool.
These constraints are checked during position creation.

\subsection{Deleveraging}

The system also includes a deleverage mechanism.
This mechanism was introduced because liquidations require the entire LP position to be withdrawn from the pool.
Since the withdrawn assets can often be used to repay a portion of the debt, it is possible that the margin level of the position increases enough to avoid imminent liquidation. Deleveraging is triggered when the position’s margin level goes below \texttt{deleverage\_margin}, which is set to a value higher than \texttt{liq\_margin} and acts as a safety buffer to help avoid liquidations.

Deleveraging is a permissioned action, meaning that third parties cannot trigger it.
It is handled by the admin system.
There is no penalty, reward, or fee taken for deleveraging, as no actual liquidation is happening: assets are just withdrawn from the pool and used to repay debt.
In case the admin system fails, the position will be liquidated when the margin level goes below \texttt{liq\_margin} by the standard (permissionless) liquidation mechanism.

\subsection{Liquidation}

Liquidation is a permissionless action that can be triggered by anyone.
Normally, if the deleveraging system is functioning correctly, the position will be fully deleveraged (i.e., all liquidity withdrawn from the pool and used to repay debt) before it reaches the liquidation margin level.
If that is not the case, deleveraging is performed by the liquidators after the position falls below the \texttt{liq\_margin} level.
If the position remains below the \texttt{liq\_margin} after that, it can be liquidated.

It is important to note that once the position is fully deleveraged, all of the collateral will be entirely in one of the two pool assets, while the debt will be in the other.
This happens because all assets withdrawn from the LP position, and any extra collateral, are used to repay as much of the debt as possible, leaving only one type of collateral and the debt in the other asset.
At this point, liquidation reduces to selling the remaining collateral to repay the debt.

A \texttt{liq\_bonus}, defined as a percentage of the liquidated debt, is taken from the liquidated position and given to the liquidator as an incentive to liquidate the position.
Liquidators are always guaranteed to receive this bonus, even if the position is below the critical margin level, in order to minimize the amount of bad debt.

\section{Conclusion}

This article formalizes and models concepts related to a concentrated liquidity protocol with generalized leverage.
The leverage is generalized in the sense that the user's assets may consist of an LP position as well as optional free collateral that consists of one or both of the LP pool's tokens.

We describe and prove multiple properties of such a system, addressing the questions posed in the introduction:
\begin{itemize}

\item The margin level function has a predictable behavior within an arbitrary price range, guaranteed by the fact that the function does not have local minima within its domain of definition.

\item Given certain reasonable assumptions, position liquidations, reductions, and deleveraging do not decrease the position's margin level, making them safe to perform.

\item Spot price manipulation cannot force unnecessary liquidations or bad debt, as long as the protocol has access to accurate oracle prices (which can be time-weighted prices from the AMM pool itself).
\end{itemize}
Additionally, we briefly examined an implementation of such a system in the Kai Leverage protocol and touched on some practical aspects.
We hope this article will be useful to DeFi practitioners, including liquidity providers, protocol designers, risk analysts, and security researchers.

\bibliographystyle{IEEEtran}

\begin{thebibliography}{1}
\providecommand{\url}[1]{#1}
\csname url@samestyle\endcsname
\providecommand{\newblock}{\relax}
\providecommand{\bibinfo}[2]{#2}
\providecommand{\BIBentrySTDinterwordspacing}{\spaceskip=0pt\relax}
\providecommand{\BIBentryALTinterwordstretchfactor}{4}
\providecommand{\BIBentryALTinterwordspacing}{\spaceskip=\fontdimen2\font plus
\BIBentryALTinterwordstretchfactor\fontdimen3\font minus
  \fontdimen4\font\relax}
\providecommand{\BIBforeignlanguage}[2]{{%
\expandafter\ifx\csname l@#1\endcsname\relax
\typeout{** WARNING: IEEEtran.bst: No hyphenation pattern has been}%
\typeout{** loaded for the language `#1'. Using the pattern for}%
\typeout{** the default language instead.}%
\else
\language=\csname l@#1\endcsname
\fi
#2}}
\providecommand{\BIBdecl}{\relax}
\BIBdecl

\bibitem{adams2021uniswap}
H.~Adams, N.~Zinsmeister, M.~Salem, R.~Keefer, and D.~Robinson, ``{Uniswap v3
  Core},'' Uniswap Labs, Tech. Rep., 2021.

\bibitem{lambert2022panoptic}
G.~Lambert and J.~Kristensen, ``Panoptic: the perpetual, oracle-free options
  protocol,'' \emph{arXiv preprint arXiv:2204.14232}, 2022.

\bibitem{RL}
M.~Romero and B.~Kalin, ``{Revert Lend Protocol},''
  \url{https://github.com/revert-finance/lend-whitepaper/blob/main/Revert_Lend-wp.pdf},
  accessed: 2024-08-19.

\bibitem{YLDR}
``What is yldr.com,'' \url{https://yldr.gitbook.io/yldr}, accessed: 2024-08-19.

\bibitem{kunalabs}
``Kuna Labs Website,'' \url{https://kunalabs.io/}, accessed: 2024-08-19.

\bibitem{liquiditymath}
A.~Elsts, ``{Liquidity Math in Uniswap v3},''
  \url{https://ssrn.com/abstract=4575232}, 2021.

\end{thebibliography}

\newpage

\section*{Appendix}

\subsection*{Proof: The margin function has no local minima}


To prove that the margin function $M(P)$ has no local minima in its domain of definition ($P>0$), we will analyze its derivative and check for the absence of negative-to-positive zero crossings within this domain.

For out-of-range positions, the margin function is monotonic, and this property is trivial. Same goes for the case when the user only borrows $Y$. Let us focus on in-range positions with non-zero liquidity and non-zero $X$ borrowed.

First, let us introduce a new variable $S$ such that $S \defeq \sqrt{P}$.
This substitution simplifies the algebra while preserving the locations of critical points. The margin function in terms of $S$ is:
\begin{align}
  M(S) &= \frac{V_{pos}(S) + x_C S^2 + y_C}{x_D S^2 + y_D}
\end{align}

Expanding $V_{pos}(S)$ for an in-range position with liquidity $L$, we have:
\begin{align}
  M(S) &= \frac{L(2 S - S_a - S^2/S_b) + x_C S^2 + y_C}{x_D S^2 + y_D}
\end{align}

This expression can be rearranged to the standard quadratic form:
\begin{align}
  M(S) &= \frac{(x_C - L / S_b) S^2 + 2 L S  + (y_C - L S_a)}{x_D S^2 + y_D}
\end{align}

This is an equation of the form:
\begin{align}\label{eq:m_general}
  M(S) &= \frac{a S^2 + b S + c}{d S^2 + e}\text{,}
\end{align}
According to our domain-specific assumptions discussed previously, the coefficients $b$ and $d$ are guaranteed to be positive, and $e$ non-negative.

The derivative of $M(S)$ with respect to $S$ is:
%
%
\begin{align}
  M'(S) = \frac{- b d S^2 + 2 (a e  - c d) S + b e}{(d S^2 + e)^2}
\end{align}
Since the denominator is positive for any non-zero $S$, the sign of $M'(S)$ depends on the sign of the numerator, which is a quadratic expression $A S^2 + B S + C$, where:
\begin{align}
  A = -b d \text{,} \quad
  B = 2 (a e - c d)  \text{,} \quad
  C = b e\label{eq:m_deriv_coeff}
\end{align}
From Eqs.~\ref{eq:m_general} and \ref{eq:m_deriv_coeff}, $A$ is always negative, while $C$ and therefore also $-AC$ always non-negative.
Let us analyze the signs of the roots $S_{1,2} = \frac{-B \pm \sqrt{D}}{2A}$, where $D = B^2 - 4AC$:
\begin{align}
  -4AC &\ge 0 \\
  B^2 - 4AC &\ge B^2 \\
  |\sqrt{B^2 - 4AC}| &\ge |B|\label{eq:b_vs_d}
\end{align}
Eq.~\ref{eq:b_vs_d} shows that $|B|$ is smaller than or equal to $|\sqrt{D}|$; it follows that $S_1$ and $S_2$ cannot both be positive.
Given that $A<0$, the quadratic equation is an inverted parabola, where $S_1$ represents a negative-to-positive crossing, which corresponds to a local minimum, and $S_2$ represents a positive-to-negative crossing, corresponding to a local maximum. However, for $M(S)$ defined in $S>0$, $S_1$ is guaranteed to lie outside its domain, leaving no local minima within the relevant range.

Uniswap v2 positions are covered by this proof as a special case, with $S_a=0$ and $S/S_b=0$, so that $V_{v2pos}(S) = 2 L S$.

\subsection*{Inverse margin level functions}

The inverse functions allow to compute the lower price $P_L$ and upper price $P_H$ where the margin function reaches a specific level $M$ for a given position $(L, p_a, p_b)$, debt $(x_D, y_D)$, and extra collateral $(x_C, y_C)$.

The functions are:
\begin{align}
  P_L(M, L, p_a, p_b, x_C, y_C, x_D, y_D) &= \frac{N_1 - 2 L \sqrt{N_2}}{D} \\
  P_U(M, L, p_a, p_b, x_C, y_C, x_D, y_D) &= \frac{N_1 + 2 L \sqrt{N_2}}{D} \,\text{,}
\end{align}
where
\begin{align*}
N_1 = &-x_C  y_C  p_b
       + x_C  y_D  M  p_b
       + x_C  L  \sqrt{p_a}  p_b \\
       &+ y_C  x_D  M  p_b
       + y_C  L  \sqrt{p_b}
       - x_D  y_D  M^2  p_b \\
       &- x_D  L  M  \sqrt{p_a}  p_b
       - y_D  L  M  \sqrt{p_b}
       - L^2  \sqrt{p_a}  \sqrt{p_b}\\
~\\
N_2 = &-x_C  y_C  p_b^2
      + x_C  y_D  M  p_b^2
      + x_C  L  \sqrt{p_a}  p_b^2 \\
      &+ y_C  x_D  M  p_b^2
      + y_C  L  p_b^{3/2}\\
      &- x_D  y_D  M^2  p_b^2
      - x_D  L  M  \sqrt{p_a} p_b^2
      - y_D L M p_b^{3/2} \\
      &- L^2 \sqrt{p_a} p_b^{3/2}
      + L^2 p_b^2\\
~\\
  D = &x_C^2 p_b
      - 2 x_C x_D M p_b
      - 2 x_C L \sqrt{p_b}
      + x_D^2 M^2 p_b
      + 2 x_D L M \sqrt{p_b} + L^2
\end{align*}

\vfill

\end{document}